\documentclass{aa}
\usepackage{graphics}
\begin{document}

  \thesaurus{03           
               (09.11.1;  
                11.09.1;  
                11.09.4;  
                11.11.1;  
                11.19.2)} 

\title{Fabry-Perot observations of the ionized gas in NGC 3938\thanks{Based 
on observations made with the William Herschel Telescope operated on 
the island of La Palma by the Isaac Newton Group in the Spanish 
Observatorio del Roque de los Muchachos of the Instituto de
Astrof\'{\i}sica de Canarias}}

\author{J. Jim\'enez-Vicente \inst{1}, E. Battaner \inst{1}, 
M. Rozas \inst{2},H. Casta\~neda \inst{2}, \and C. Porcel \inst{1}}

\institute{Departamento de F{\'\i}sica Te\'orica y del
 Cosmos, Universidad de Granada, E-18071, Granada, Spain
\and Instituto de Astrof{\'\i}sica de Canarias. E-38200, La Laguna. 
Tenerife. Spain}

\date{Received; accepted}
\titlerunning{F-P observations of the ionized gas in NGC 3938}
\authorrunning{J. Jim\'enez-Vicente et al.}
\maketitle
\begin{abstract}

  The nearly face-on spiral galaxy NGC 3938 has been observed in
  the $H_\alpha$ line with the TAURUS
  II Fabry-Perot interferometer at the William Herschel Telescope in
  order to study the kinematics of the ionized
  gas. We are able to construct intensity, velocity and velocity
  dispersion maps for this galaxy. The rotation curve of the galaxy 
is calculated up to 4.5 radial scale lengths from
  the galactic centre. The residual velocity field shows very small
  values with no systematic
  pattern. The mean velocity dispersion is approximately constant with
  radius at about 11 km/s as previously reported for the neutral and 
molecular gas. We have also studied the relation between intensity and
velocity dispersion for the ionized gas. We have found that this
  distribution is compatible with a turbulent gas relaxing to a
  Kolmogorov type turbulence as the stationary regime. The average
  dispersion varies with intensity as $\sigma \propto I^{1/8}$
  although it becomes much steeper at higher intensities, for which
  the dispersion is kept almost constant at a value of about
19 km/s.


\end{abstract}

\section{Introduction}

Kinematical studies of spiral galaxies are often performed by means of HI
observations. This allows a study of the global kinematics (and gas
distribution) throughout the whole galaxy up to very large radii,
though it requires large observation times. As an alternative
complementary technique, Fabry-Perot
interferometry allows a very large amount of information to be
extracted with quite short observation times. Although many studies
should be carried out with the neutral atomic gas,
Fabry-Perot interferometry may be a very good alternative if we are
interested in the global kinematics of the galaxy. On the other hand,
as we are observing a different emisson mechanism, differences in the
deduced dynamical properties (i.e. in the neutral and ionized gas
dynamics) may be a clue to understand some galactic kinematical
features. We can extract
standard information such as velocity maps, rotation curves, etc with very
good spatial and spectral resolution.

In the case of face-on galaxies, Fabry-Perot interferometry is
very adequate for the study of vertical motions. We can calculate
residual velocity maps (which allow to study large scale vertical motions),
velocity dispersions (whose knowledge is essential to understand the
vertical equilibrium in the galaxy (see Combes \& Bequaert, 1997))
etc. 
In particular it is specially
well suited to study vertical motions over star forming regions
which can be the key to understand disk-halo interactions according to
some theoretical models (Norman \& Ikeuchi, 1988). Even at larger
scales, random motions, are strongly related with the magnetic 
difussivity in the disk, which could be a key ingredient in
understanding the origin and structure of the galactic magnetic field.

Along this paper we extensively extract these kind of data for the
galaxy NGC 3938.
In Section 2 we review general properties of the galaxy. Section 3 is
devoted to the observational procedure and data reduction. In Section
4 we discuss the distribution of the $H_\alpha$ emission. The
kinematics is analysed in Section 5, including the rotation curve
calculation and velocity dispersion analyses. The relation between
intensity and velocity dispersion is studied in Section 6. The main
results and conclusions are finally summarized in Section 7.

\section{NGC 3938}
\begin{figure}[tbp]
\resizebox{8.8cm}{!}{\includegraphics{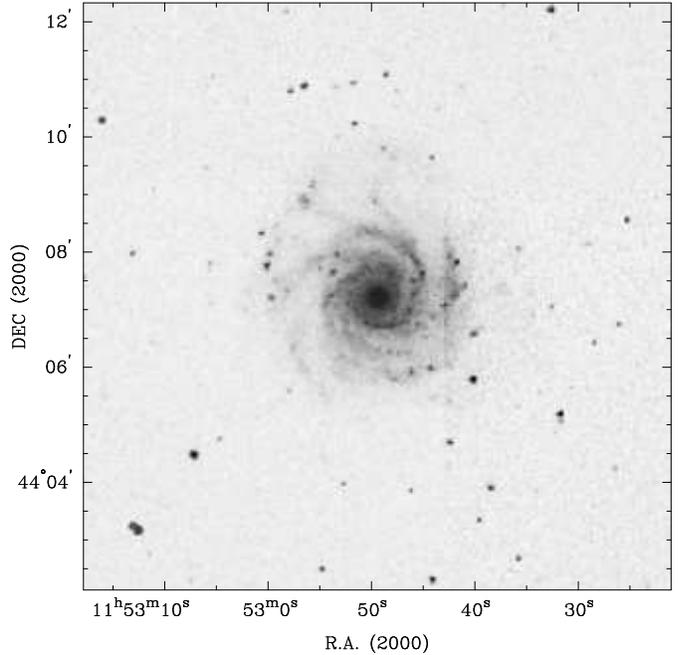}}
\caption{Optical image of NGC 3938}
\label{image1}
\end{figure}

NGC 3938 is a nearly face-on (inclination $\approx 14^\circ$) late type (Sc) 
spiral galaxy (see figure \ref{image1}\footnote{
Based on photographic data of the National Geographic Society -- Palomar
Observatory Sky Survey (NGS-POSS) obtained using the Oschin Telescope on
Palomar Mountain.  The NGS-POSS was funded by a grant from the National 
Geographic Society to the California Institute of Technology.  The      
plates were processed into the present compressed digital form with     
their permission.  The Digitized Sky Survey was produced at the Space   
Telescope Science Institute under US Government grant NAG
W-2166.}). Table \ref{tab:prop} summarizes the main galaxy
properties. It
has quite well defined arms and is rich in HII regions. Its
distance derived from its corrected radial velocity (848 km/s
according to the RC3 catalog) is
11.3 Mpc (using a Hubble constant of 75 km/s/Mpc). At this distance,
we have 54.78 pc/arcsec, or 61.36 pc/pixel, and therefore we are able
to construct high resolution maps. The galaxy belongs to the Ursa
Major Cluster (see Tully et al. 1996) but has no close companions
within 100 kpc. Its radial scale length is 31.1 arcsec or 1.7 kpc
(this value is for the K band as measured by Tully et al. 1996,
although they found a clear trend to increase with decreasing
wavelength). van der Kruit \& Shostak (1982) carried out surface photometry of
this galaxy in the F band and they fitted an exponential disc with a
scale length of 36 arcsec, but they found a faster decline at $R
\approx 2.5'$. It is approximately up to this point where our
observations are able to reach. 

The stellar velocity dispersion was measured by Bottema (1988). He
stated that the central vertical stellar velocity dispersion is 
low with a value of about 30
km/s, and about 15 km/s at one radial scale length, compatible with an
exponential decline as found in other galaxies (see Bottema (1993)),
although he warns that in the outer region the values are below the
resolution limit. 
For the bulge he found a velocity dispersion of about 40 km/s.

This galaxy has been observed in HI
by van der Kruit \& Shostak (1982) at Westerbork with a $24''\times
36''$ beam. They found no systematic pattern of z-motions in excess of
5 km/s, although in a subsequent analysis of the residual velocity map
by Foster \& Nelson (1985) they found a ring and a spoke pattern. The
velocity dispersion was observed to be constant with radius at a value
of 10 km/s. CO observations with the IRAM 30-m telescope
by Combes \& Becquaert along the major and minor axis also found this
behaviour for the molecular gas with a slightly smaller velocity
dispersion of 8.5 km/s.

\begin{table}
\caption{Galaxy properties}
  \label{tab:prop}
  \begin{tabular}{ll}
    \hline \\
    Name             & NGC 3938 \\
    Type             & Sc(s)I \\
    R.A.(2000)       & $11^h52^m49.8^s$ \\
    Dec (2000)       & $44^\circ 07'26''$ \\
    B Magnitude      & 10.9 \\
    $D_{25}$         & $5.4'$ \\
    $V_\odot$ (RC3)  & 809 km/s \\
    $V_{gsr}$ (RC3)    & 848 km/s \\
    Distance         & 11.3 Mpc \\
    Redshifted $H_\alpha$ & 6580.5\AA \\
    Inclination      & $14^\circ$ \\
    P.A.             & $20^\circ$ \\
    Environment      & Ursa Major Cluster. \\
                     & No close ($<100$kpc) companions \\
    \hline
   \end{tabular}
\end{table}

The HII region luminosity function has been calculated by McCall et
al. (1996) by two methods (according to their nomenclature): Fixed
Threshold Photometry (FTP) and Percentage of Peak Photometry
(PPP). The luminosity function slope above the completeness limit
(which they take $L_t=38.9$ ergs/s and $L_t=38.3$ ergs/s for each
 method respectively)
is identical in both cases and its value is $1.7\pm 0.1$. This value is
within the range reported by Kennicutt et
al. (1989) and Rozas et al. (1996) in other galaxies. It is worth
noting that this completeness limit is quite high as compared with the
values used by other authors such as Kennicutt et al. (1989) and Rozas et
al. (1996). This value is much closer to the observed turnover in the
luminosity function observed by Kennicutt et al. (1989) for galaxies
of the same type and which he attributes to a transition between
normal giant regions and supergiant regions. The same effect has been
observed by Rozas et al. (1996) at a value of 38.6 ergs/s, and they
explain this effect as a transition between ionization bounded and
density bounded HII regions. If the completeness limit used by McCall
et al (1996) is taken as such a turnover, then the slope is low when
compared  with reported values for this range of the luminosity
function in other galaxies.

\section{Observations and data reduction}
The observations were made on March 18, 1997, at the William
Herschel Telescope. The night was not photometric and the seeing
conditions were not very good with a measured seeing of 3 arcsec. This
avoided photometric calibration of our data and therefore we will use
arbitrary units for intensity throughout this paper.
The galaxy was observed in the $H_\alpha$ line with the TAURUS II
Fabry-Perot interferometer at the Cassegrain focus of the William
Herschel Telescope. We used the $500 \mu$
etalon and the f/2.11 camera with a TEK CCD detector. This setup gives a
field of view of about 5 arcmin (because the interference filter at the
focal plane is vignetting the field) and a pixel size of 0.56 arcsec. 
We rebinned 2x2 the
CCD reading which results in a pixel size of 1.12 arcsec. The nominal
spectral resolution of the etalon at 6500 \AA\  is 9.75 km/s with a free
spectral range of 195 km/s. We scanned the spectral range in 55
steps with an exposure time of 120 seconds each, giving a total of
about 2 hours to take the full datacube. A calibration cube with a
CuNe lamp was
taken at the beginning of the night, and calibration rings were taken
before and after the object exposure in order to make the phase
calibration. The measured instrumental profile obtained after phase 
correcting the calibration datacube was 7.7 km/s, and the spectral
resolution was 3.61 km/sec/pixel. Table \ref{tab:setup} summarizes the
observational setup.

\begin{table}
\caption{Observational parameters}
  \label{tab:setup}
  \begin{tabular}{ll}
    \hline \\
    Date of observation & 18/03/1997 \\
    Telescope  & WHT \\
    Focus      & Cassegrain \\
    Instrument & TAURUS II \\
    Etalon     & $500 \mu$ \\
    Filter     & 6577/15 \\
    Detector   & TEK CCD \\
    Steps      & 55 \\
    Exposure  time per step  & 120 sec \\
    Free Spectral Range & 194.1 km/sec \\
    Instrumental width & 7.7 km/s \\
    Spectral resolution  & 3.61 km/sec/pixel \\
    Pixel size   &  1.12 arcsec \\
    Seeing       &  3 arcsec \\
    \hline
   \end{tabular}
\end{table}

The observed datacube is bias substracted and phase corrected by
standard procedures.
We used the 6577/15
filter for order sorting. This filter allows the OH night sky line at
6577.28 \AA\ to reach the etalon (at a different interference order) and
therefore this spectrum is added to the target one. This effect should
be corrected before the datacube is analysed. To do the correction,
large regions of the observed field are chosen at places where there 
is no emission
from the target and a spectrum is calculated for each of those regions. An
average spectrum is calculated from all of them and an
artificial datacube is created for a homogeneous source with that
spectral content which is then substracted from the original one.

We have also created lower resolution (12 arcsec) datacubes by
smoothing the original one with a wider beamsize in order to have
smoother maps as usually done with radio data. The rest of the data
reduction procedure is the same as for the full resolution datacube,
for which we briefly describe it bellow.

To extract the physical information from the phase corrected datacube
we used the GIPSY package. First of all, we substract the continuum from
the full datacube using the velocity channels that have no $H_\alpha$
line emission, and then each spectrum is fitted to a gaussian
function, from which we extract the velocity (line mean position),
velocity dispersion (line width) and intensity (proportional to the
line peak intensity) at each position. We have also used a standard
moments procedure to calculate these maps. The velocity maps obtained 
by both methods match very well, but not so the
dispersion map for which we prefer the gaussian fitting method. 
The reason to choose gaussian
fitting instead of a moments procedure is to allow a better
determination of velocity dispersions and avoid a systematic bias, 
 as pointed out by van der Kruit
\& Shostak (1982). 

The calculated maps are very noisy. To clean them we put constraints
on the intensity and velocity dispersion selecting only those pixels
for which these parameters exceed a certain value. Finally, some
remaining noise is eliminated by hand.
The calculated velocity dispersion should be corrected because of the
instrumental width (which is 7.7 km/s) and the natural line width
(which is 3 km/s for the $H_\alpha$ line). 
These numbers should be quadratically subtracted
from the measured dispersion. As we are mainly interested in calculating the 
non-thermal
velocity dispersion, we should also correct for the thermal width. We
have taken a typical temperature of  $10^4 K$ for the ionized gas
(Spitzer, 1978 and Osterbrock, 1989). At this temperature the thermal 
width is 9.1 km/s.

Therefore we have, finally
\begin{equation}
  \label{dispcor}
  \sigma_{nt}=(\sigma_{obs}^2-\sigma_{inst}^2-\sigma_{nat}^2
-\sigma_{th}^2)^{1/2}
\end{equation}
where $\sigma_{nt}$, $\sigma_{obs}$, $\sigma_{inst}$ and $\sigma_{th}$
are the non-thermal, observed, instrumental and thermal components 
respectively.

\section{The ionized gas distribution}

\begin{figure}
\resizebox{8.8cm}{!}{\includegraphics{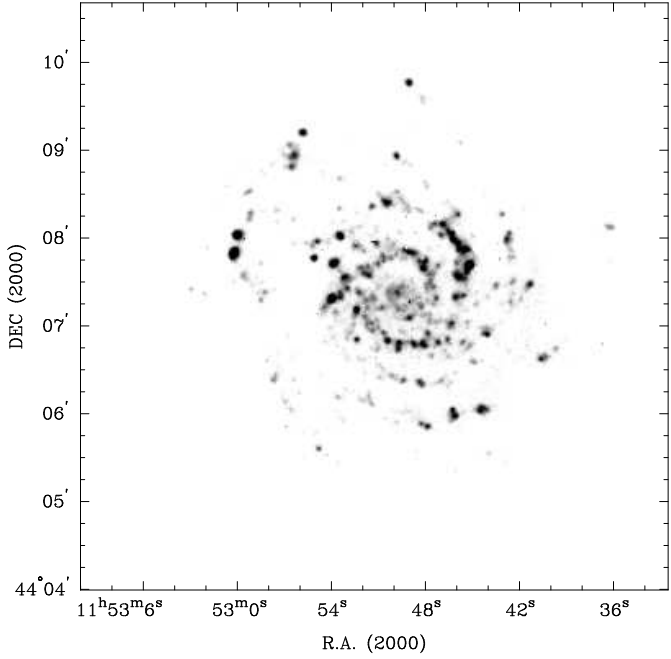}}
\caption{Intensity map for NGC 3938}
\label{inten1}
\end{figure}
\begin{figure}
\resizebox{8.8cm}{!}{\includegraphics{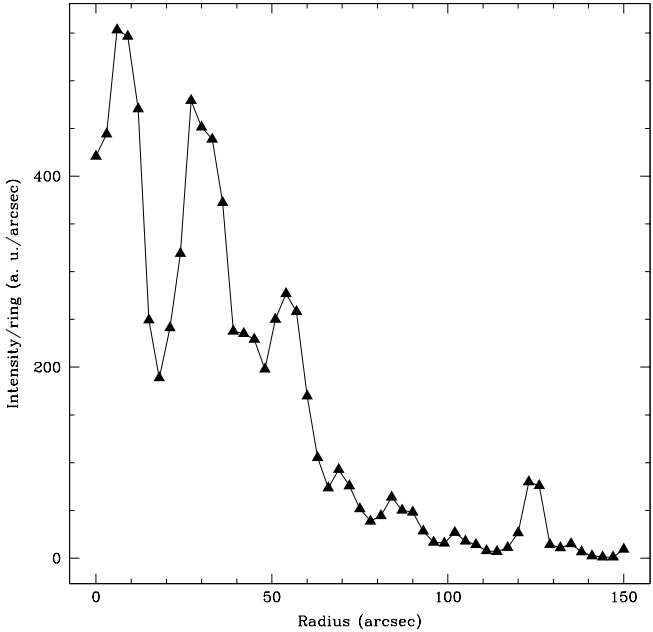}}
\caption{Radial distribution of the ionized gas in NGC 3938}
\label{inten2}
\end{figure}

The spatial distribution of the ionized gas can be seen in figure
\ref{inten1}. Most of it comes from the HII regions but there is also
emission coming from regions which can not be clearly identified as
HII regions. They can be low luminosity HII regions or belong to a 
{\em diffuse} component. Down to our detection level, most of this low 
intensity emission is seen
in the periphery of HII regions. The HII regions are highly
concentrated in the arms, mostly in the inner disc, but there are also
giant HII regions at large galactocentric radii such as the two bright
regions in the east. The northern arm is brighter than the southern
one, and most of the bright HII regions are hosted by it. This will
affect the global line profile as we will see later.
 
The radial distribution of the ionized gas can be seen in figure
\ref{inten2}. It has been calculated by adding the intensity over
rings with a width of 3.36 arcsec and dividing by the ring area.
The peaks at about 30 and 50 arcsec from the center are due to the spiral
arms (specially at about 30 arcsec from the center, the spiral arms
resemble a ring) and the peak at about 125 arcsec is due mainly to the pair of
supergiant HII regions in the east.

\section{Kinematics}

\begin{figure*}
\resizebox{17.8cm}{!}{\includegraphics{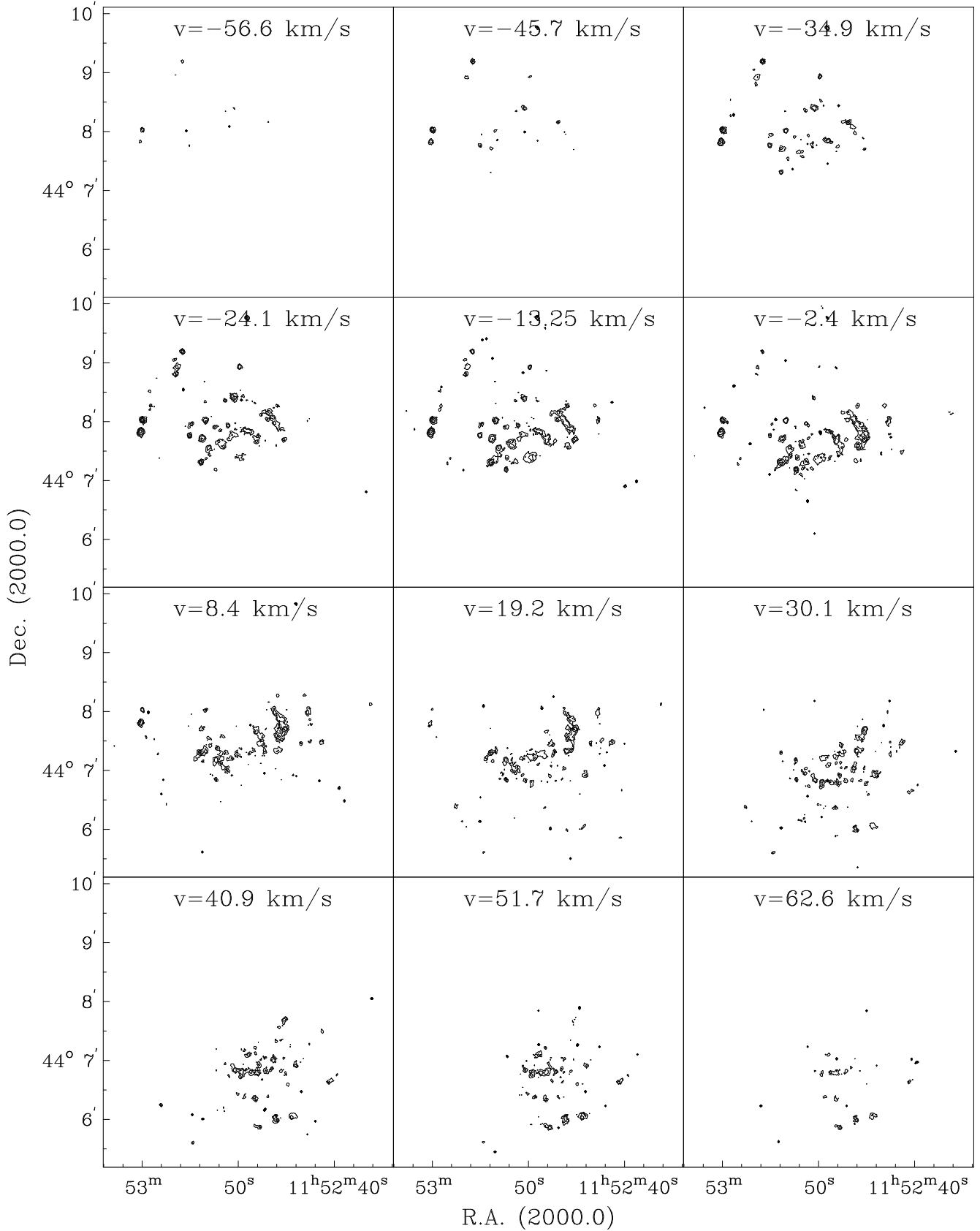}}
\caption{Channel maps of the continuum substracted $H_\alpha$ emission
  for NGC 3938 at 1.12 arcsec/pixel resolution. The velocity of each
  channel (with the systemic velocity of the galaxy substracted) is
  shown at the top of each panel.}
\label{panel}
\end{figure*}

The full resolution (1.12 arcsec/pixel) channel maps for the
$H_\alpha$ line emission are shown in figure \ref{panel}. Only each
third channel is shown (with a separation of 10.83 km/s between
them). Only the brightest emission regions are shown to avoid noise.
A first inspection shows no special features but normal rotation and the
absence of warping.

\begin{figure}
\resizebox{8.8cm}{!}{\includegraphics{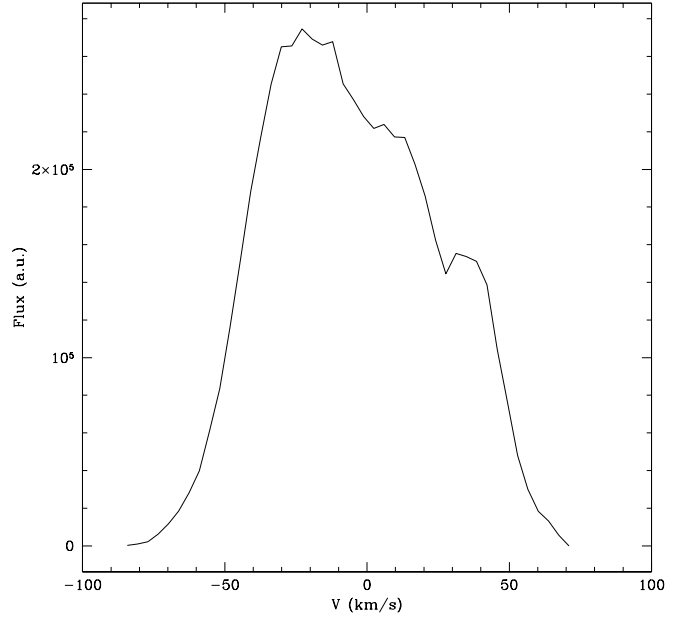}}
\caption{Integrated $H_\alpha$ profile of NGC 3938}
\label{global}
\end{figure}

We have calculated the integrated $H_\alpha$ profile by summing the
intensity in each channel. It is shown in figure \ref{global}.
The profile does not show the usual double horn structure that can be
seen in the HI profile for this galaxy in van der Kruit \& Shostak
(1982) and in other galaxies. This is partially due because the
central part of the disc which is usually devoid of neutral atomic gas
has now a quite large amount of emission. This is most often the case
with $H_\alpha$ profiles and another clear example can be seen in
Fig. 5 of Amram et al. (1996) with the galaxy DC 1842-63 n$^\circ$
24. This region appears in
channels close to the systemic velocity and therefore these channels 
show a larger flux than observed for the neutral gas as compared with
the channels in the horns. On the other hand the profile asymmetry
which can also be seen partially in the HI profile is due to the
larger emission of the northern arm (as can be seen in figure
\ref{global}) which is very rich in bright HII regions and which is 
in the receding part of the galaxy. 

In the rest of this section we will study in detail the velocity and
dispersion maps as well as the calculation of the rotation curve.

\subsection{The velocity map}

\begin{figure}[tbp]
\resizebox{8.8cm}{!}{\includegraphics{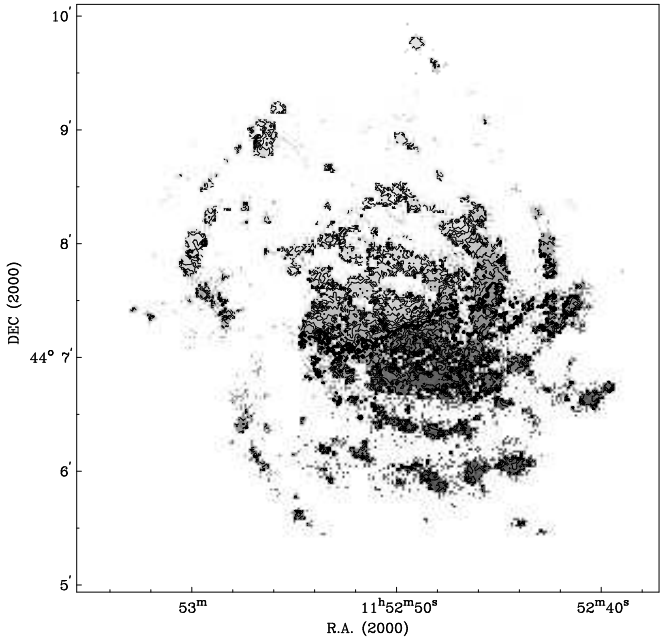}}
\caption{High resolution (1.12 arcsec/pix) velocity map for NGC
  3938. Levels are shown each 5 km/s. Dark gray shows the
  approaching region while light gray shows the receding region. 
The thick line shows the 0 km/s line.}
\label{vmap1}
\end{figure}

\begin{figure}[tbp]
\resizebox{8.8cm}{!}{\includegraphics{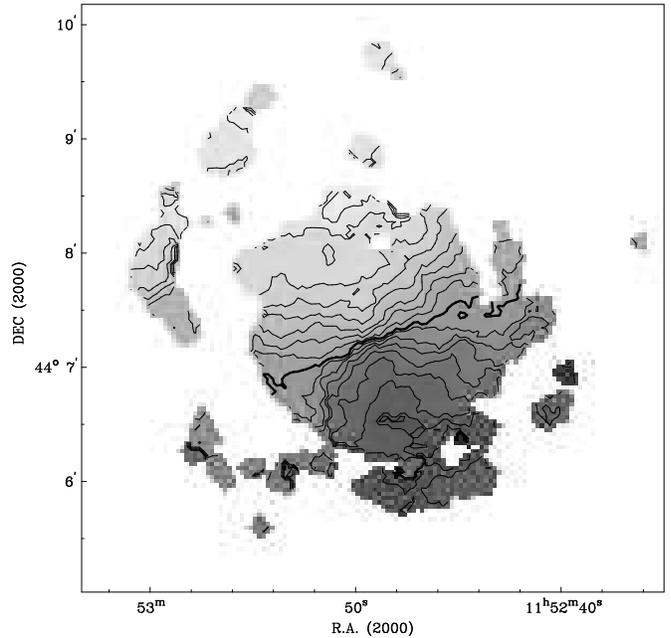}}
\caption{Velocity map for NGC 3938 at 12 arcsec resolution. Levels are
  shown each 5 km/s. Dark gray shows the
  approaching region while light gray shows the receding region.
The thick line shows the 0 km/s line.} 
\label{vmap2}
\end{figure}

The velocity map at the highest available resolution (1.21
arcsec/pixel) is shown in figure \ref{vmap1} as calculated with the
moments procedure. The global kinematics is better seen in the
smoother map at 12 arcsec resolution shown in figure \ref{vmap2}. It
shows no special features but normal rotation although there are small
traces of streaming motions near the inner arms. Anyway, these
deviations from axisymmetric rotation are very small as we will see
later. There
are not traces of a warp as can be seen in the kinematical major axis
which, is kept almost straight up to the largest radii. Our
observations do not reach up to very large galactocentric radii and
the warp could start further and be hidden to the present observations, but
this is not very likely in this case as no warp has been observed
in HI for which there are available data up to
much larger radii.

\subsection{Rotation curve}

\begin{figure}[tbp]
\resizebox{8.8cm}{!}{\includegraphics{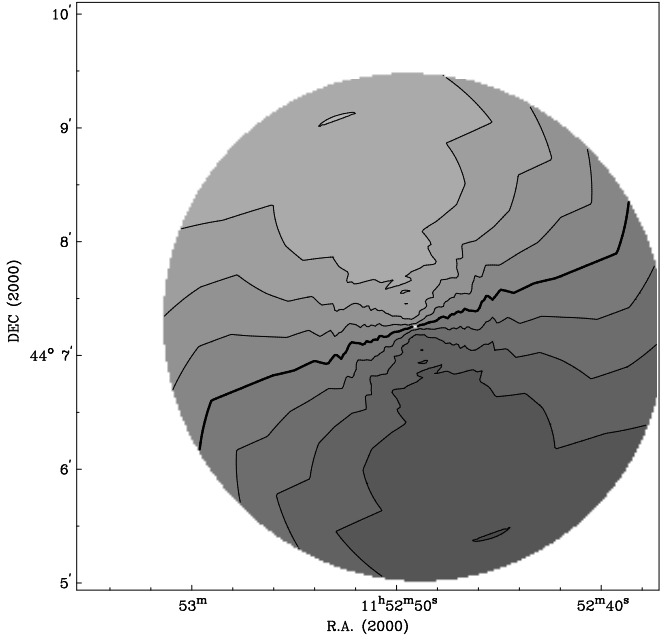}}
\caption{Model velocity map for NGC 3938 calculated by fitting a
  tilted rings model to the observed velocity map.  Levels are shown 
each 10 km/s.
Dark gray shows the
  approaching region while light gray shows the receding region. 
The thick line shows the 0 km/s line.}
\label{model}
\end{figure}
\begin{figure}[tbp]
\resizebox{8.8cm}{!}{\includegraphics{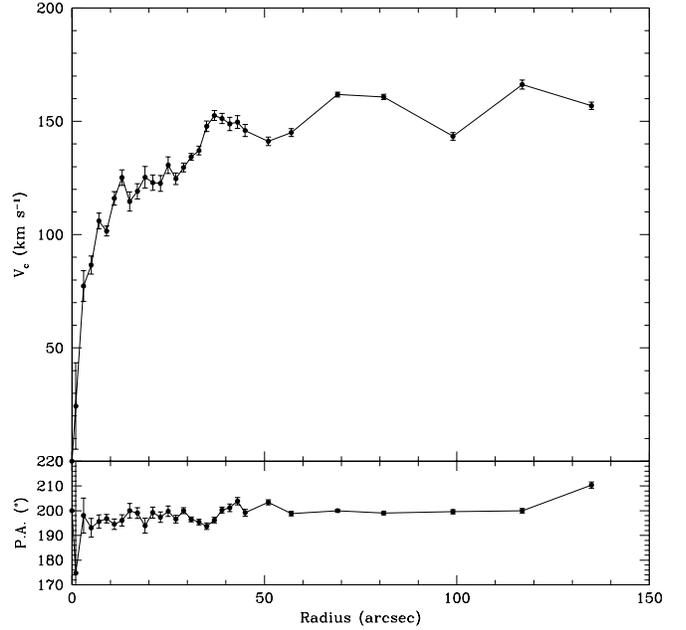}}
\caption{Rotation curve for NGC 3938 (upper panel) and fitted
  position angle (lower panel). Error bars
  represent least-squares fit errors. The points were derived from the
  high resolution map but taking wider rings for the outer parts.}
\label{rotcur}
\end{figure}
\begin{figure}[tbp]
\resizebox{8.8cm}{!}{\includegraphics{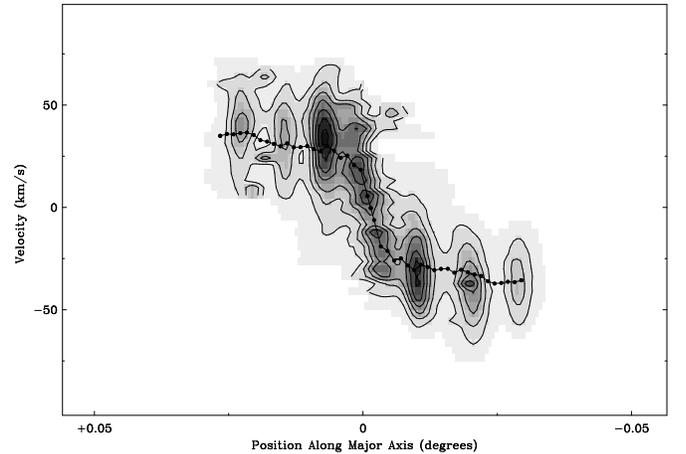}}
\caption{Position-velocity diagram of NGC 3938 along the major axis
  (P. A.=$20^\circ$) with the inner part of the calculated rotation
  curve superimposed.}
\label{slice}
\end{figure}

We have used the high resolution (1.12 arcsec/pix) velocity field
using a tilted rings model fitting as described by Begeman (1989)
using the ROTCUR task of the GIPSY package. The
galaxy is divided into concentric rings and each of them is
characterized by its inclination angle ($i$), position angle of the
major axis (P.A.) and rotational velocity ($V_c$). In principle
additional parameters can be fitted, such as the centre of each ring and its
systemic velocity. These are too many parameters for a single fit and
we preferred to calculate the latter at an earlier stage and
then fix them in the final fitting procedure. Each data point in the
velocity map is weighted by $|\cos (\theta)|$ where $\theta$ is the
azimuthal angle measured from the major axis. Moreover, data with
$|\theta < 5^\circ |$ are excluded from the fit. 

The low inclination angle of NGC 3938 makes it difficult to fit
simultaneously the rotational velocity and the inclination
angle. As this galaxy shows no signs of a warp (see the velocity map
in van der Kruit \& Shostak (1982)) it is safe to use a constant
inclination throughout the whole disc. We have chosen a value of
$i=14^\circ$. The effect of using a different angle but close to this
value would be either to
raise (if it were larger) or to lower (if it were smaller) the value of
the rotational velocity, but its shape would remain unchanged.

As the number of data points in the outer part of the disk is very
low, we have used wider rings in this region in order to keep a
minimum number of pixels in each ring. The width of the rings has been
calculated by demanding a minimum of 300 data points inside each
ring. Therefore the outer parts of
the fitted model are of poorer resolution, and so is the rotation curve.

The modeled velocity map for NGC 3938 is shown in figure
\ref{model}. The rotation curve of the galaxy can be plotted with the
calculated values for the rotational velocity of each ring and is
shown in figure \ref{rotcur} with the fitted position angle of each ring.

This is the best determination of the rotation curve of NGC 3938 up to
date. The value of the rotational velocity at large galactocentric
radii agrees with previously published values at about 
$V_{rot}=38\csc i$ km/s (see van der Kruit \& Shostak, 1982 and Combes
\& Becquaert, 1997), but our determination of the curve has a much 
larger spatial
resolution, specially at small radii. Bottema (1993) using the
Tully-Fisher relation deduced a value of $V_{rot}=147\pm 20$ km/s
(corresponding to an inclination of $15^\circ$) which is compatible
with our determination of 157 km/s. At larger radii, the details of
the shape of the rotation curve are not so reliable because of the
small number of data points in the fit, but it gives the right
value of the maximum rotational velocity if we assume it to be
constant at large radii.

Figure \ref{slice} shows a position-velocity diagram along the major
axis with the inner part of the calculated rotation curve
superimposed. It can be seen that the shape of the rotation curve
matches very well the gas distribution. 

\subsection{The residual velocity map}

\begin{figure}[tbp]
\resizebox{8.8cm}{!}{\includegraphics{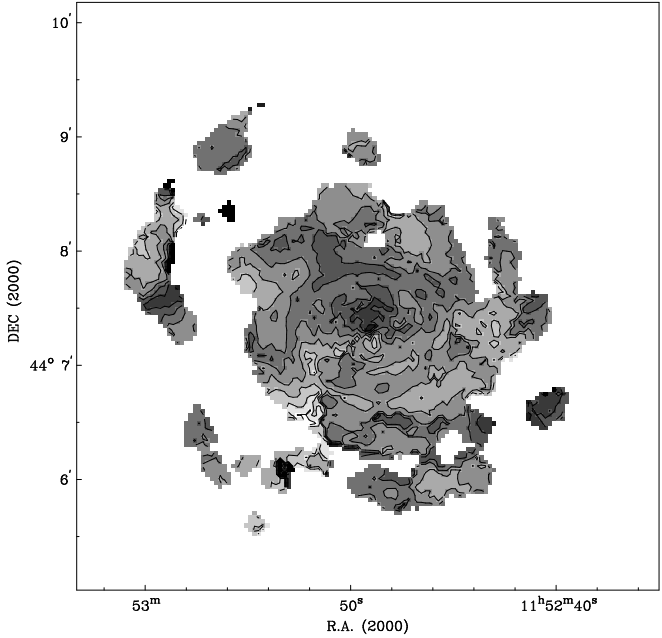}}
\caption{Residual velocity map for NGC 3938. Contour levels are
  between -12 and 12 km/s (from light gray to dark gray) separated by 3 km/sec each. }
\label{res}
\end{figure}

The residual map at 12 arcsec resolution calculated by subtracting
the fitted model velocity map from the observed map is shown in figure
\ref{res}.  The rms residual velocity calculated with the 12 arcsec
resolution residual map is 4.6 km/s. We do not detect any systematic 
pattern in excess 10 km/s
although some traces of the spiral arms can be seen in the inner
parts. We have tried to compare this residual map with the one
published by Foster \& Nelson (1985). We find no similarities between
them and although we have not repeated their logarithmic spiral
decomposition it is not very likely that the ring and the spoke are
detectable in our map.

Some regions with a noticeable residual velocity seem to be associated with
bright HII regions with very high velocity dispersions.This is, for
instance, the case for the
northward one in the pair of regions in the furthest east which shows a
systematic defect of about 10 km/s. There 
is no clear evidence
that this residual is a true movement of these regions but the
coincidence is worth being noted. Phenomena such as the chimneys proposed by
Norman and Ikeuchi (1989) would show this behavior, and filaments of
$H_\alpha$ emission leaving the disk from bright HII regions have also
been detected in edge on galaxies (see for example Dettmar
(1990)). These chimneys are associated with the most active star
forming regions as the OB associations hosted by supergiant HII
regions, and are a very interesting possibility in understanding the
disc-halo interaction in galaxies. Further
observational research would be desirable to clarify this point.

\subsection{The velocity dispersion map}

\begin{figure}[tbp]
\resizebox{8.8cm}{!}{\includegraphics{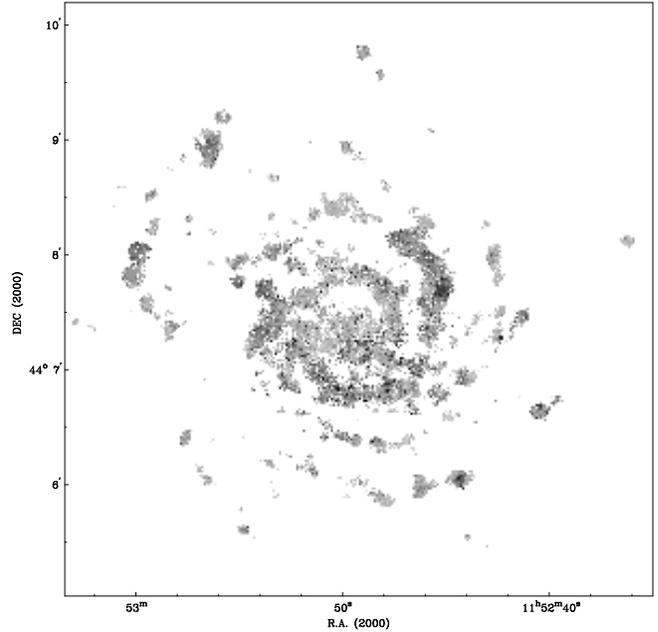}}
\caption{Velocity dispersion map at 1.12 arcsec/pixel
  resolution. Gray levels range from 0 to 20 km/s (from light gray to
  dark gray). }
\label{dismap}
\end{figure}

\begin{figure}[tbp]
\resizebox{8.8cm}{!}{\includegraphics{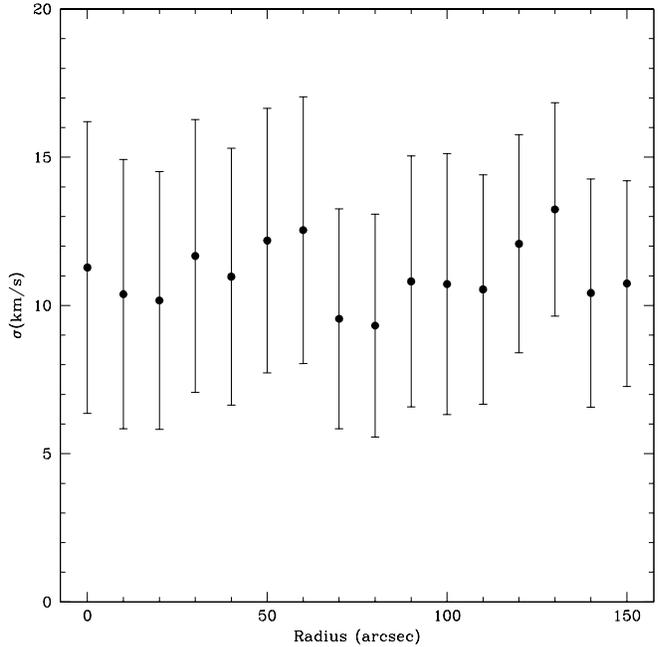}}
\caption{Radial distribution of velocity dispersion in rings of width
11 arcsec. The data have been corrected for instrumental, natural and
thermal broadening.}
\label{disvsrad}
\end{figure}

        We calculated the velocity dispersion map by the gaussian
fitting method as commented upon in section 3. The map is shown in
figure \ref{dismap}. Some of the brightest HII regions are also
visible in this map by their high velocity dispersion which goes up
to 17 km/s. Anyway, there is no clear relation between intensity and
velocity dispersion as we will see later. 
The mean velocity dispersion of the $H_\alpha$ emission
in the whole galaxy is about 11 km/s. This value is observed to be constant
within the observational errors at every galactocentric radii as
shown in figure \ref{disvsrad}. This effect had already been found for
the neutral gas by van der Kruit \& Shostak (1982) and for the
molecular gas by Combes \& Bequaert (1997). These latter authors discuss
in their paper several possible explanations for this coincidence in
the vertical velocity dispersions of the different gas
phases. Moreover, they found that the velocity dispersion is higher than
the value needed to maintain gravitational stability. Stellar
formation is rejected by them as the
heating source because most of the HI showing this behavior is well
outside the optical disc. Of course this is not the case for the
ionized gas where new born stars are the most important heating
mechanism, and which show a very similar value for the velocity
dispersion. Therefore, some combination of several heating mechanisms
could be acting to produce the observed behavior.

\section{Intensity versus velocity dispersion}

\begin{figure}
\resizebox{8.8cm}{!}{\includegraphics{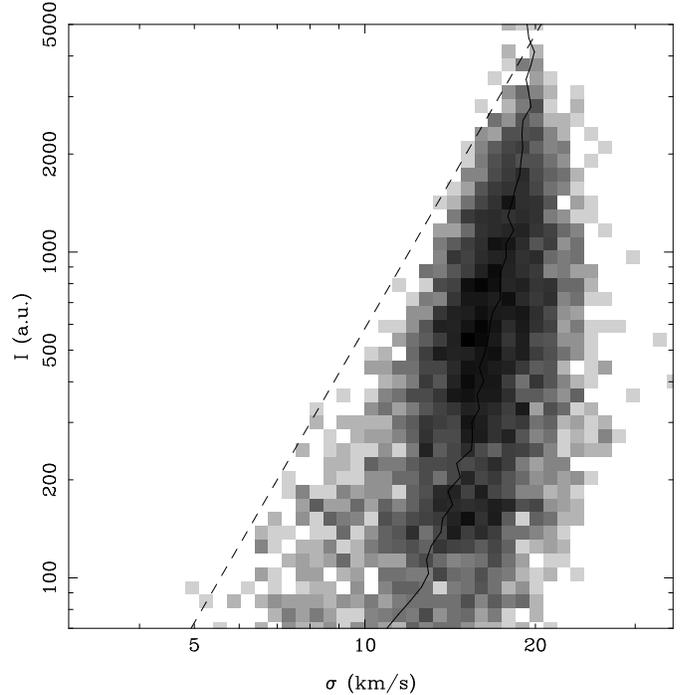}}
\caption{Velocity dispersion vs intensity distribution. Solid line
represents the mean dispersion at each intensity. Dashed line
represents a function $I\propto \sigma^3$ as the upper envelope.}
\label{disvsint}
\end{figure}

We have studied the relation between velocity dispersion and
intensity in NGC 3938. In this purpose we have constructed the
two-dimensional distribution shown in figure \ref{disvsint} where we
show the velocity dispersion (uncorrected) vs intensity for each
pixel. We are aware that the region of low dispersions (below the
instrumental width) in
the histogram is not fully reliable, but we include it for
completeness. We have
calculated the mean dispersion at a given intensity (shown with the
solid line in figure \ref{disvsint}). The mean dispersion grows in
average as
$\sigma \propto I^{1/8}$, although it becomes much flatter for high
intensities, for which the dispersion becomes almost constant at a
value of about 19 km/sec. Another interesting feature is the upper
envelope of the distribution shown by the dashed line. The line is a
function $I \propto \sigma^3$. A relation ${\cal L} \propto \sigma^3$ has
been previously found for HII regions (note that our distribution is
built for every pixel and not for HII regions) by Roy et
al. (1986). We have tested if this also happens in this 
galaxy, but we
have not found any correlation between dispersion and luminosity in a
sample of 25 HII regions of NGC 3938. Arsenault et al. (1990) studied
such a correlation between luminosity and dispersion in NGC 4321 and
did not find any correlation if they used the full HII regions sample,
although they found a correlation if they used just the brightest HII
regions. We cannot confirm nor reject such a correlation with our data.

We have tried to interpret the upper envelope in the $(I-\sigma)$
distribution in terms of Kolmogorov turbulence. In this case, energy
losses per volume unit and time unit would be $\rho \sigma^3 /L$,
being $\rho$ the density and $L$ the size of the HII region. The
turbulent energy production (i. e. $\partial /\partial t (1/3 \rho
\sigma^2)$) would be of type $xQ$ where $Q$ is the energy density
input due to stars and $x$ a proportionality constant with a very
small value. When thermal and turbulent energies reach stationary
conditions, then, if production terms are proportional for both, so
must be their losses. Thermal energy density losses due to radiation would be
per time unit $\sigma_{SB} T^4 /L$ (where $\sigma_{SB}$ is the
Stefan-Boltzmann constant, and $T$ is the temperature). 
Under stationary conditions one would have $x
\sigma_{SB} T^4 = \sigma^3 \rho$. The intensity is $I
\propto \sigma_{SB} T^4 l^2$ (where $l$ is a length corresponding to
the pixel size) and we would
have thus $I\propto \sigma^3$. In figure \ref{disvsint} we see that
the upper envelope matches this relation such that true dispersion is
always larger or equal than predicted by that relation. This could be
understood if violent stellar phenomena (such as strong stellar winds of
SN explosions) produce very large turbulent motions which afterwards
trend to
relax to a stationary regime compatible with Kolmogorov turbulence. 
This is also partially confirmed by some HII regions showing high
dispersions in their outer envelopes which suggest expansive motions
which could have been produced by shocks resulting from SN explosions or
strong stellar winds inside them.

There is an interesting case which is the previously commented
supergiant HII region (northward of the pair to the east) with a 
systematic defect in the residual
velocity. Its outer shell shows a large velocity dispersion (even
larger than in the center of the HII region) with low intensity. This
could be interpreted as an expanding shell. In this case the outer
shell high velocity dispersion could be interpreted in terms of high
temperature (of about $5\times 10^4 K$) due to shock heating. Anyway
we warn that our spatial resolution is at the limit to detect this
feature, and therefore higher
spatial and spectral resolution observations could clarify this point.

\section{Results and Conclusions}

        We have studied the distribution and kinematics of the ionized
gas in the spiral galaxy NGC 3938 from Fabry-Perot observations. Our
main results and conclusions are:

\begin{itemize}
\item The velocity map of the galaxy is compatible with pure axisymmetric
  rotation. There are no traces of a warp and very small traces of 
streaming motions across the spiral arms.

\item The rotation curve of the galaxy shows a standard shape with an
asymptotic rotation velocity of about 157 km/s (for an inclination
angle of $14^\circ$). The rotation curve is compatible with previous
determinations with HI and molecular gas observations, but we are able to
determine it with much better spatial resolution in the inner 2 arcmin
of the galaxy.

\item The velocity dispersion of the ionized gas is kept constant with
  galactocentric
  radius at a value of about 11 km/s. This value is similar (although
  slightly higher) to previously reported values for the neutral and
  molecular gas in this galaxy. Also we find that there is no radial
  dependence of the vertical velocity dispersion.

\item We find that the velocity dispersion vs intensity distribution has
  an upper envelope of the type $I\propto \sigma^3$, compatible with a
  Kolmogorov turbulence regime as the stationary regime. The average
  intensity vs velocity dispersion relation is much steeper with a
  mean behavior of the type $I\propto \sigma^8$ or even steeper for
  higher intensities, for which the dispersion is kept almost constant
  at a value of about 19 km/s.

\end{itemize}
\begin{acknowledgements}
We are very grateful to the referee Dr. M. Marcelin for his valuable
  comments and remarks on this paper.
  This paper has been supported by the spanish
  ``Ministerio de Educaci\'on y Cultura'' (PB96-1428) and the ``Plan
  Andaluz de Investigaci\'on'' (FQM-0108).
\end{acknowledgements}

\end{document}